\documentclass[aps,prl,reprint,showpacs,floatfix]{revtex4-1}

\usepackage{amsmath}
\usepackage{slashed}
\usepackage{txfonts}
\usepackage{microtype}
\usepackage{graphicx}
\usepackage[breaklinks=true]{hyperref}
\usepackage{color}

\def\sout{\bgroup\markoverwith
{\textcolor{red}{\rule[0.5ex]{2pt}{0.5pt}}}\ULon}
\def\be{\begin{equation}}
\def\ee{\end{equation}}
\def\bes{\begin{equation*}}
\def\ees{\end{equation*}}
\def\bea{\begin{eqnarray}}
\def\eea{\end{eqnarray}}
\def\beas{\begin{eqnarray*}}
\def\eeas{\end{eqnarray*}}
\def\bal#1\eal{\begin{align}#1\end{align}}
\def\bals#1\eals{\begin{align*}#1\end{align*}}
\newcommand{\bra}[1]{\langle #1|}
\newcommand{\ket}[1]{|#1\rangle}
\newcommand{\braket}[2]{\langle #1|#2\rangle}

\renewcommand{\vec}{\vectorsym}
\newcommand{\del}{\partial}
\renewcommand*{\vec}[1]{\boldsymbol{#1}}

\graphicspath{{figures/}}

\usepackage[normalem]{ulem}

\begin{document}

\title{Emergence of non-abelian magnetic monopoles in a quantum impurity problem}

\author{E. Yakaboylu}
\email{enderalp.yakaboylu@ist.ac.at}
\author{A. Deuchert}
\email{andreas.deuchert@ist.ac.at}
\author{M. Lemeshko}
\email{mikhail.lemeshko@ist.ac.at}

\affiliation{IST Austria (Institute of Science and Technology Austria), Am Campus 1, 3400 Klosterneuburg, Austria}

\date{\today}

\begin{abstract}

 Recently it was shown that molecules rotating in superfluid helium can be described in terms of the angulon quasiparticles [Phys. Rev. Lett. \textbf{118}, 095301 (2017)]. Here we demonstrate that in the experimentally realized regime the angulon can be seen as a point charge on a 2-sphere interacting with a gauge field of a non-abelian magnetic monopole.  Unlike in several other settings, the gauge fields of the angulon problem emerge in the real coordinate space, as opposed to the momentum space or some effective parameter space. Furthermore, we find a topological transition associated with making the monopole abelian, which takes place in the vicinity of the previously reported angulon instabilities. These results pave the way for studying topological phenomena in experiments on molecules trapped in superfluid helium nanodroplets,  as well as on other realizations of orbital  impurity problems.
\end{abstract}

\maketitle

In Maxwell's unification of electricity and magnetism, there was one piece missing that would make the electric and magnetic forces perfectly symmetric with respect to each other -- the  magnetic monopoles. As demonstrated by Dirac~\cite{Dirac_1931}, the existence of a single magnetic monopole would explain quantisation of electric charge everywhere in the Universe. Since Dirac's work, the existence of magnetic monopoles -- as real elementary particles or effective quasiparticles -- has preoccupied physicists working in  several different fields. In high-energy physics,  't~Hooft~\cite{Thooft_1974} and Polyakov~\cite{Polyakov_1974} demonstrated the existence of non-abelian magnetic monopoles in the context of the unification of the fundamental interactions~\footnote{see Preskill's notes~\cite{Preskill_1984} for a detailed review, and the references therein}.
Despite  the lack of experimental evidence for \textit{elementary} monopoles in nature~\cite{milton2006theoretical},  collective phenomena exhibiting the  behavior of  magnetic monopoles have been predicted to emerge in condensed matter systems~\cite{castelnovo2008magnetic, chuang1991cosmology,PhysRevLett.102.080403,PhysRevLett.103.030401,
PhysRevA.84.063627}, and subsequently observed in experiment~\cite{fang2003anomalous,
milde2013unwinding,morris2009dirac, ray2014observation}.

As  shown by Berry~\cite{berry1984quantal}, magnetic monopoles can also emerge in an external parameter space of a simple quantum mechanical problem~\cite{wilczek1984appearance,Aharonov_1987}. Moreover, the parameter space can be  further generalized to coordinates of a particle interacting with another particle~\cite{Moody_86,wilczek1989geometric,bohm1992berry}. For example, Moody~\textit{et~al.}~\cite{Moody_86} showed that the effective Hamiltonian 
describing nuclear rotation of a diatomic molecule can be rewritten as that of a charged particle interacting with a gauge field of a magnetic monopole. In follow-up studies the emerging gauge fields have been studied in various contexts, from different types of atomic problems~\cite{zygelman1990non,RevModPhys.83.1523,PhysRevLett.110.170402,ruseckas2005non} to spinor Bose-Fermi mixtures \cite{phuc2015controlling}, to the fractional quantum Hall effect~\cite{PhysRevLett.57.1195}, to Bose-Einstein Condensates~\cite{PhysRevLett.102.080403,PhysRevLett.103.030401,
PhysRevA.84.063627}. Finally, emergence of monopole-like gauge fields represents an important tool for computing topological invariants of quantum systems using Chern numbers~\cite{nakahara2003geometry}, and thereby classifying the topology of the problem~\cite{PhysRevLett.97.036808,PhysRevLett.117.015301,
PhysRevLett.113.050402,morimoto2016topological,PhysRevLett.102.080403}. Such a topological classification of quantum states is particularly relevant in the context of current research on topological states of matter~\cite{PhysRevLett.49.405,PhysRevB.31.3372,
PhysRevLett.95.146802,PhysRevLett.98.106803,PhysRevB.76.045302,PhysRevLett.61.2015}. 

In this Letter, we demonstrate that non-abelian magnetic monopole fields emerge in the recently introduced angulon impurity problem~\cite{Lemeshko_2015, PhysRevX.6.011012, Lemeshko_2016_book}. The angulon represents a quantum impurity exchanging orbital angular momentum with a many-particle bath, and serves as a reliable model for the rotation of molecules in superfluids~\cite{lemeshko2016quasiparticle, YuliaPhysics17, Shepperson16, Shepperson17}. We hereby show that  superfluid helium nanodroplets, which have been used as a tool of molecular spectroscopy for over two decades~\cite{toennies2004superfluid,toennies1998spectroscopy,callegari2001helium,
choi2006infrared,stienkemeier2006spectroscopy}, behave as effective non-abelian magnetic monopoles with respect to molecular  impurities trapped inside them. Furthermore, our analysis  reveals a topological transition taking place around the previously reported `angulon instabilities'~\cite{Cherepanov, Lemeshko_2015, Yakaboylu16}, where orbital angular momentum is resonantly transferred between the impurity and the bath. Such a transition corresponds to an abelianization of the magnetic monopole (making the monopole abelian).

Our approach is based on  the idea of Moody~\textit{et al.}~\cite{Moody_86}, which we herein extend to the case of quantum impurity problems, involving an infinite number of degrees of freedom. Let us start from the most general Hamiltonian describing a quantum impurity interacting with a many-particle environment:
\be
\label{qp_ham}
\hat{H} (\vec{r}) = - \mu \, \vec{\nabla}^2 + \hat{H}_\text{mb}(\vec{r}) \, .
\ee
 Here the Laplacian in the generalized coordinates of the impurity, $\vec{r}$, represents the kinetic energy of the impurity.  For a linearly-moving impurity corresponding to the polaron problem~\cite{Devreese15}, there is $- \mu \, \vec{\nabla}^2 \equiv 1/(2m) \hat{\vec{P}}^2$, where $m$ is the mass of an electron moving at momentum $\vec{P}$  (units of $\hbar \equiv 1$ are used hereafter). In the angulon problem, for a rotating impurity we have $- \mu \, \vec{\nabla}^2 \equiv B \hat{\vec{L}}^2 $, where $\hat{\vec{L}}$ is the angular momentum operator and $B = 1/(2I)$ the rotational constant with $I$ the moment of inertia. The second term of Eq.~\eqref{qp_ham}, corresponds to the many-body part of the Hamiltonian, which includes the kinetic energy of the bath and the impurity-bath interactions that depend on $\vec{r}$. Furthermore, $\hat{H}_\text{mb}(\vec{r})$ can include any external potential such as that due to an electromagnetic field~\cite{PhysRevB.31.3689, Redchenko16, Lemeshko_2013}.

The eigenvalue equation for Hamiltonian~(\ref{qp_ham}) can be written as
\be
\label{eigenvalue_eqn}
\hat{H} (\vec{r}) \ket{\Psi^{\alpha} (\vec{r})} = E^\alpha \ket{\Psi^{\alpha} (\vec{r})} \, ,
\ee
where $\ket{\Psi^{\alpha} (\vec{r})} \equiv \braket{\vec{r}}{\Psi^{\alpha}}$ is the eigenstate in the coordinate space of the impurity, and $\alpha$ is the quantum number labeling the eigenstate. For example, in the case of the angulon, $\alpha$ corresponds to  the total orbital angular momentum of the entire system, $L$, and its projection on the quantization axis, $M$~\cite{Lemeshko_2015}. 

Next, we perform the Born-Oppenheimer (BO) expansion of  the eigenstates~\cite{Moody_86,wilczek1989geometric,bohm1992berry} :
\be
\label{BO_expansion}
\ket{\Psi^{\alpha}  (\vec{r})} = \sum_n \Phi_n^{\alpha}  (\vec{r}) \ket{\varphi_n (\vec{r})} \, ,
\ee
where $n$ can label any possible quantum numbers, $\ket{\varphi_n (\vec{r})}$ are the basis vectors formed from the eigenstates of $\hat{H}_\text{mb}(\vec{r})$ in the corresponding Fock space, and $\Phi_n^{\alpha}  (\vec{r}) \equiv \braket{\varphi_n (\vec{r})}{\Psi^\alpha (\vec{r})}$. After we plug the BO expansion~(\ref{BO_expansion}) into Eq.~(\ref{eigenvalue_eqn}), and project onto the basis vector $\bra{\varphi_m (\vec{r})}$, we obtain the following equation
\be
\label{eigenvalue_eff_hamil}
\sum_n H_{mn} (\vec{r}) \Phi_n^{\alpha}  (\vec{r}) =  E^\alpha \Phi_m^{\alpha}  (\vec{r}) \, ,
\ee
with the effective impurity Hamiltonian 
\be
\label{gauge_cov_qp}
H_{mn}(\vec{r}) =  - \mu \sum_l \vec{D}_{ml} \cdot \vec{D}_{ln} + A^0_{mn} (\vec{r}) \, .
\ee
Here $A^0_{mn} (\vec{r}) = \bra{\varphi_m (\vec{r})} \hat{H}_\text{mb}(\vec{r}) \ket{\varphi_n (\vec{r})} $ is the non-abelian scalar potential, and $\vec{D}_{mn} \equiv \vec{\nabla}\delta_{mn} - i \vec{A}_{mn} (\vec{r})$ is the covariant derivative. The particular object that we are interested in is the non-abelian gauge field
\be
\label{gauge_field}
\vec{A}_{mn} (\vec{r}) = \bra{\varphi_m(\vec{r})} i \vec{\nabla} \ket{\varphi_n (\vec{r})}  \, , 
\ee
which contains all the information about the geometry and topology of the problem.

Thus, we have rewritten the impurity Hamiltonian~(\ref{qp_ham}) in the gauge covariant form~(\ref{gauge_cov_qp}) corresponding to the gauge group $U(\infty)$. Note that no approximations have been introduced during this step.  The origin of the emerging gauge symmetry, or more properly the gauge redundancy~\cite{seiberg2006emergent}, follows from the fact that one could  unitarily transform the basis vectors, $\ket{\varphi_{n'} (\vec{r})} \to \ket{\varphi'_{n'} (\vec{r})} =   U_{n n'}  (\vec{r}) \ket{\varphi_n (\vec{r})} $ (summation is implied over the repeating index),  such that the BO expansion~(\ref{BO_expansion}) be invariant under the transformation $\Phi_{n'}^{\alpha}  (\vec{r}) \to {\Phi'}_{n'}^{\alpha}  (\vec{r}) = U^\dagger_{n' n} (\vec{r}) \Phi_n^{\alpha}  (\vec{r})$. In fact, the latter transformation, together with
\be
\label{gauge_trans}
\vec{A}'_{n' m'} (\vec{r}) = U^\dagger_{n' n} (\vec{r}) \vec{A}_{n m} (\vec{r}) U_{m m'}(\vec{r}) + i U^\dagger_{n' n} (\vec{r})\vec{\nabla} U_{n m'}  (\vec{r}),
\ee
leaves the Schr{\"o}dinger equation~(\ref{eigenvalue_eff_hamil}) invariant.

Now, we assume that there exists a certain physical configuration where the restriction of the basis vectors is legitimate. A standard technique would be to use the product-state ansatz (the BO approximation), where the eigenstate~(\ref{BO_expansion}) is approximated by $\ket{\Psi^{\alpha}  (\vec{r})} \approx \Phi_n^{\alpha}  (\vec{r}) \ket{\varphi_n (\vec{r})}$, which results in a $U(1)$ gauge field. This approximation is, in principle, applicable to any impurity problem. A non-abelian gauge field, on the other hand, can only be obtained by considering more than one many-body state. The latter can be realized, for instance, within the adiabatic approximation. There it is assumed that the many-body state $\ket{\varphi_n (\vec{r})}$ remains in a certain energy level, and an $N$-fold degenerate levels yields a $U(N)$ gauge field. In many-body systems, however, the latter can be challenging to achieve unless the many-body state of interest is separated from the rest of the spectrum by an energy gap. Here, in order to truncate the number of basis vectors, we apply the variational principle to the Hamiltonian $H (\vec{r})$. As long as the quasiparticle has a discrete energy spectrum (as the angulon does), the variational state can be written as the BO expansion with a small number of basis vectors. We note that the discussion below is applicable to other quantum impurity problems with discrete spectrum, such as a polaron interacting with a magnetic field~\cite{PhysRevB.31.3689}, or a particle in a double-well potential coupled to a bosonic bath~\cite{RevModPhys.59.1}.

Let us consider the angulon quasiparticle as a particular example~\cite{Lemeshko_2015, Lemeshko_2016_book, PhysRevX.6.011012, Bikash16, Redchenko16, Li16, Yakaboylu_2017}. For the angulon, the truncation of basis states through the variational principle was shown to provide a good agreement with experiment~\cite{lemeshko2016quasiparticle,  Shepperson16}.  The angulon Hamiltonian, originally derived in Ref.~\citep{Lemeshko_2015} is given by:
\bal
\label{angulon_ham}
& \hat{H}^\text{A} (\hat{\Omega}) = B \vec{L}^2 \\
\nonumber & + \sum_{k \lambda \mu} \omega (k) \, \hat{b}^\dagger_{k \lambda \mu} \hat{b}_{k \lambda \mu} + \sum_{k \lambda \mu} U_{\lambda} (k) \left[ Y^{*}_{\lambda \mu}(\hat{\Omega}) \hat{b}^\dagger_{k \lambda \mu} + Y_{\lambda \mu}(\hat{\Omega}) \hat{b}_{k \lambda \mu}  \right] \, .
\eal
Here $\sum_k \equiv \int d k$, $\omega (k) $ is  the dispersion relation for the bosonic bath, and $\hat{b}^\dagger_{k \lambda \mu} $ and $\hat{b}_{k \lambda \mu}$ are the bosonic creation and annihilation operators. For the latter we use the angular momentum representation, with $k$, $\lambda$, and $\mu$ labeling the bosonic linear momentum, angular momentum, and its projection on the laboratory-frame $z$-axis, respectively (see Ref.~\cite{Lemeshko_2016_book} for more details). The last term of Eq.~\eqref{angulon_ham} describes the interaction of the impurity with the bosonic bath, where $Y_{\lambda \mu}(\hat{\Omega})$ are the spherical harmonic operators~\cite{Varshalovich} that depend on the impurity orientation in the laboratory frame, $\hat \Omega \equiv (\hat \theta, \hat \phi)$, and $U_{\lambda} (k)$ is the angular-momentum-dependent coupling strength. 

In Ref.~\cite{PhysRevX.6.011012}, it was shown that in the strong-coupling regime, $U_\lambda \gg B$, the  angulon can be described using the following variational state
\be 
\label{variational_state}
\ket{\Psi^{LM}} = \hat{S}_1 \hat{S}_2 \left[ g_{0} \ket{0} \ket{LM0}  +    \sum_{k \lambda n} \alpha_{k \lambda n}  \hat{b}^\dagger_{k \lambda n} \ket{0} \ket{LM n} \right] \, ,
\ee 
where  $g_{0}$ and $\alpha_{k \lambda n}$ are the variational parameters, and $n$ labels the projection of total angular momentum $\vec{L}$ on the body-fixed quantisation axis~\cite{LevebvreBrionField2}. The first transformation,
\begin{equation}
\hat{S}_1 = e^{-i \hat \phi \otimes \hat \Lambda_z}~e^{-i \hat \theta \otimes \hat \Lambda_y}~e^{-i \hat \gamma \otimes  \hat  \Lambda_z},
\end{equation}
 brings the bath degrees of freedom into the frame co-rotating along with the quantum rotor, where $  \vec{\hat  \Lambda} = \sum_{k \lambda \mu \nu} \vec{\sigma}^\lambda_{\mu \nu} \hat{b}^\dagger_{k \lambda \mu} \hat{b}_{k \lambda \nu} $ is the total angular momentum operator of the bosonic bath and $\vec{\sigma}^\lambda$ the $\lambda$'s representation of the rotation group. The second transformation,
\begin{equation}
\hat{S}_2 = \exp \left[ - \sum_{k \lambda} f_\lambda (k) \left( \hat{b}^\dagger_{k \lambda 0} - \hat{b}_{k \lambda 0} \right) \right],
\end{equation}
 on the other hand, is the coherent state transformation with $f_\lambda (k) = U_\lambda (k) \sqrt{(2\lambda +1)/(4 \pi)}/ [ \omega (k) + B \lambda (\lambda +1)]$. A detailed discussion can be found in Ref.~\cite{PhysRevX.6.011012}.

Following Eq.~\eqref{BO_expansion}, we can perform the BO expansion of the impurity eigenstates $\ket{\Psi^{LM} (\vec{r})}$ in the coordinate space of the impurity, i.e., with $\vec{r} \equiv \Omega$, with the following basis vectors:
\be
\ket{\varphi_n (\Omega)} = \hat{S}_1(\Omega) \hat{S}_2 \frac{1}{\sqrt{c_n}}\left( \sum_{k \lambda} \beta_{k \lambda n}  \hat{b}^\dagger_{k \lambda n} \ket{0} + \delta_{n0} \ket{0} \right) \, ,
\ee
with $\beta_{k \lambda n} \equiv \alpha_{k \lambda n}/ g_0$, and $c_n = \sum_{k \lambda} |\beta_{k \lambda n}|^2 + \delta_{n0}$. 
Then, the gauge field can be obtained from Eq.~(\ref{gauge_field}):
\bal
\vec{A}_{n m} (\Omega)  & = \frac{1}{\sqrt{c_n c_m}}\sum_{k \lambda} \left[ \beta_{k \lambda n}^{*} \beta_{k \lambda m} \vec{\Sigma}^\lambda_{n m} (\Omega) \right. \\
\nonumber & \left. - f_\lambda(k) \left( \beta_{k \lambda n}^{*} \delta_{m0} \vec{\Sigma}^\lambda_{n 0} (\Omega)  + \beta_{k \lambda m} \delta_{n0} \vec{\Sigma}^\lambda_{0 m} (\Omega)  \right) \right] \, ,
\eal
where  $\vec{\Sigma}^\lambda_{n m}(\Omega) = \left[ \sigma^{z \, \lambda}_{n m} \cos \theta - \sigma^{x \, \lambda}_{n m} \sin \theta \right] \hat{\phi} + \sigma^{y \, \lambda}_{n m}\, \hat{\theta} $. The dimension of the gauge group is determined by $2 \lambda_\text{max} + 1$, where $\lambda_\text{max}$ is the maximum angular momentum number for which the impurity-boson interaction is of relevant strength.

\begin{figure}
  \centering
  \includegraphics[width=\linewidth]{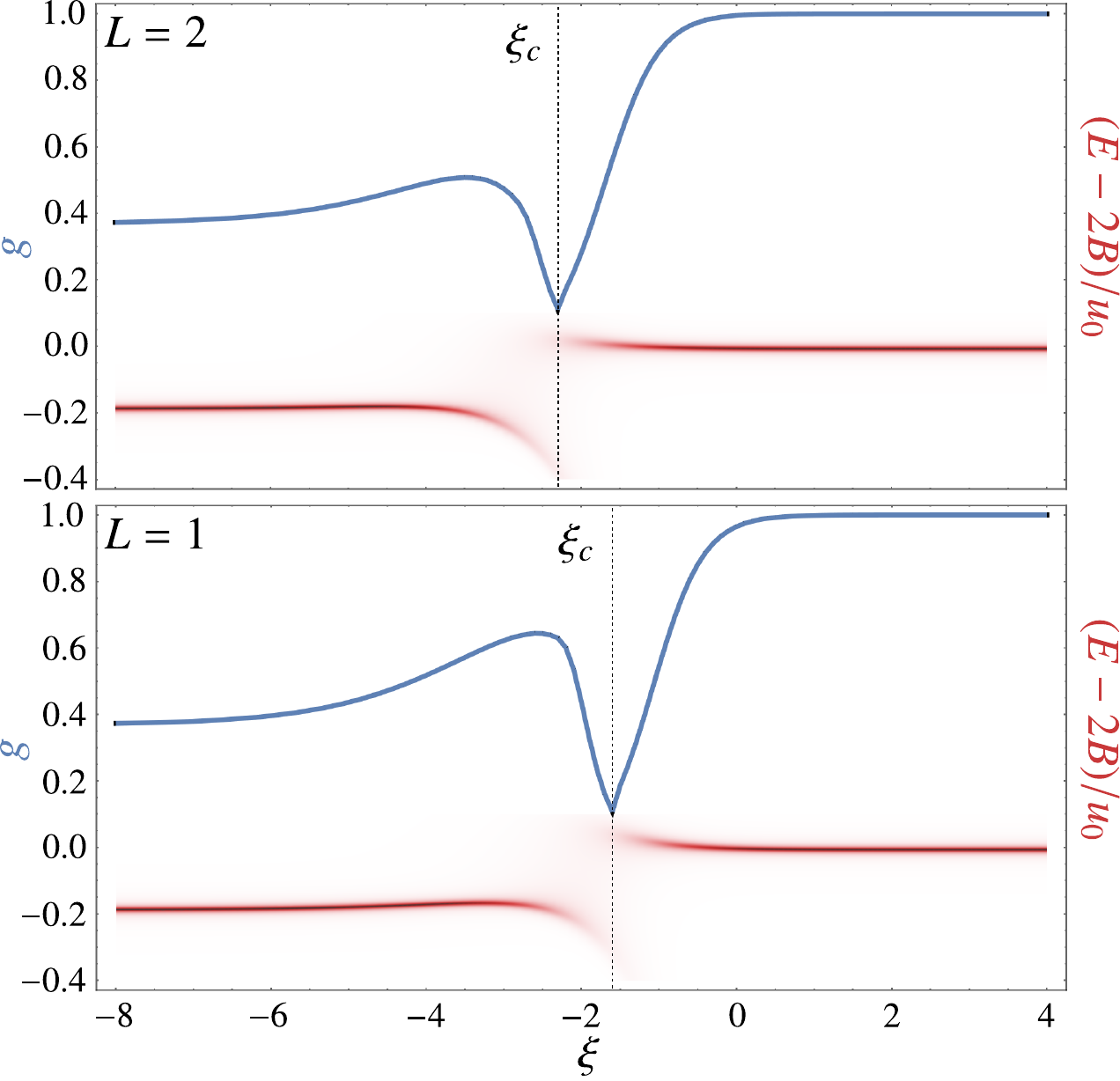}
 \caption{The angulon spectral function (red) and the charge of the $U(3)$ magnetic monopole (blue) as a function of the dimensionless rotational constant, $\xi = \ln [B/u_0]$, for $L=1$ and $L=2$ states. The dimensionless bath density is set to $\tilde{n}_0 = \ln [0.014 ]$. Vertical dashed lines indicate the critical value, $\xi_c$. See text.}
 \label{monopole}
\end{figure}

For impurities of experimental interest (such as molecules in superfluids), only a few coupling constants $U_\lambda (k)$ are of substantial magnitude~\cite{StoneBook13}. As an example, let us assume that only the isotropic term, $U_0 (k)$, as well as the leading anizotropic term, $U_1 (k)$, are present.
In this case, the gauge group of interest is $U(3)$. The explicit calculation for the gauge fields results
\be
\label{u3_field}
A_\phi = \begin{pmatrix}
- \cos \theta & \frac{-\kappa}{\sqrt{2}} \sin \theta & 0 \\
\frac{-\kappa^*}{\sqrt{2}} \sin \theta & 0 & \frac{-\kappa^*}{\sqrt{2}} \sin \theta \\
0 & \frac{-\kappa}{\sqrt{2}} \sin \theta & \cos \theta
\end{pmatrix}\, , \, A_\theta = \begin{pmatrix}
0 & \frac{i \kappa}{\sqrt{2}} & 0 \\
\frac{-i \kappa^*}{\sqrt{2}}& 0 & \frac{i \kappa^*}{\sqrt{2}}\\
0 & \frac{-i \kappa}{\sqrt{2}}  & 0
\end{pmatrix} \, ,
\ee
where $\kappa  = \sum_k \beta_{k11}^{*} (\beta_{k10}-f_1 (k))/\sqrt{c_1 c_0}$~\footnote{Since $[\Lambda_z, \hat{S}_2] = 0$ and $\exp\left(-i \gamma \Lambda_z\right) \hat{b}^\dagger_{k \lambda n} {\ket{0}} = \exp\left(-i \gamma n \right) \hat{b}^\dagger_{k \lambda n} {\ket{0}} $, we know that $\exp( - i \gamma n) D^{L \, *}_{Mn} (\Omega,\gamma) = D^{L \, *}_{Mn} (\Omega,0) $ as well as that the eigenstate ${\ket{\Psi^{LM} (\Omega)}}$ is independent from $\gamma$. Thus, we can set $\gamma=0$ without loss of generality. For this choice, however, the gauge field~(\ref{u3_field}) cannot be defined at the poles. If one chooses $\gamma = + (-) \phi$, the gauge field can also be defined on the north (south) pole, however not on both poles, which is a property of magnetic monopoles~\cite{Dirac_1931,Wu_Yang_1975}.}. We further calculate the curvature, which is given by
\bal
\nonumber F_{\phi \theta} & = i [D_\phi, D_\theta ] = \del_\phi A_{\theta}  - \del_\theta A_{\phi}  - i [A_\phi , A_\theta] \\
\label{curvature}
& =  (1- |\kappa|^2) \sin \theta \, \sigma^1_z \, .
\eal
By definition, $F_{\phi \theta}$ in Eq.~\eqref{curvature} is the strength of a $U(3)$ magnetic monopole with charge
\begin{equation}
\label{eq:g}
g= 1 - |\kappa|^2.
\end{equation}
This allows us to interpret the angulon (or a molecule immersed into a droplet of superfluid $^4$He) as an impurity interacting with the field of a non-abelian magnetic monopole.

In order to provide quantitative results, we  set the model parameters to the values used in Ref.~\cite{PhysRevX.6.011012}.  The angular-momentum-dependent coupling strength is taken to be of the following form~\cite{Lemeshko_2015}: $U_{\lambda} (k) = u_\lambda \sqrt{8 k^2 \varepsilon(k) n_0 / \left[\omega (k) (2 \lambda + 1) \right]} \int dr \, r^2 v_\lambda (r) j_\lambda (k r)$, where  $u_\lambda$ and $v_\lambda (r)$ definine the strength and shape of the molecule-boson interaction potential and $j_\lambda (kr)$ is the spherical Bessel function. We model the two-body potentials using Gaussian form factors $v_\lambda(r) = (2 \pi)^{-3/2} e^{-r^2/(2 r_\lambda^2)}$ and adapt a Bogoliubov-type dispersion relation $\omega(k) = \sqrt{\varepsilon (k)(\varepsilon(k) + 2 g_\text{bb} n_0)}$, where $\varepsilon(k) = k^2/(2m)$ is the boson kinetic energy. The boson-boson contact interaction is set to $g_\text{bb} = 418 (m^2 u_0)^{-1/2}$. Furthermore, we take the potential anisotropy to be $u_1 = 5 u_0$ and the range to be $r_0 = r_1 = 15 (mu_0)^{-1/2}$, with $u_{\lambda} \equiv 0$ for $\lambda > 1$. In what follows, we study the behavior of the system as a function of the dimensionless rotational constant, $\xi = \ln[B/u_0]$, and the dimensionless density,  $\tilde{n}_0 = \ln[n_0 (m u_0)^{-3/2}]$.

In Refs.~\cite{Lemeshko_2015, PhysRevX.6.011012, Lemeshko_2016_book} it was shown that the variational calculation with a state of the form~\eqref{variational_state} allows one to derive the Dyson equation for the angulon's Green's function~$G_L(E)$. As a result, one acquires access to the entire spectrum of the system through the spectral function, $A_L (E) = \text{Im}[G_L (E + i 0^{+})]$~\cite{AltlandSimons}. We calculate the angulon spectral function exactly in the same way as described in Refs.~\cite{Lemeshko_2015, PhysRevX.6.011012, Lemeshko_2016_book}. Fig.~\ref{monopole} shows the spectral function for $L=1$ and $L=2$ states at the density $\tilde{n}_0 = \ln[0.014]$ (red color), which corresponds to the case presented in Ref.~\cite{PhysRevX.6.011012}. One can see that around $\xi_c \approx -1.6 $ for $L=1$ and $\xi_c \approx -2.3 $ for $L=2$ there is a  discontinuity in the spectrum, which corresponds to the so-called `angulon instability'~\cite{Lemeshko_2015, PhysRevX.6.011012}. Such instabilities correspond to a resonant transfer of angular momentum between the impurity and the bath. Recently, the angulon instabilities were identified in experimentally observed spectra of CH$_3$ and NH$_3$ molecules trapped in superfluid helium nanodroplets~\cite{Cherepanov}. 

In the same figure, we present the strength of the magnetic-monopole $g$, Eq.~\eqref{eq:g}. First of all,  we observe that the monopole strength $g$ approaches  zero around the instability point $\xi_c$, which corresponds to the limit of  $\kappa \to 1$. In this limit the magnetic field strength~(\ref{curvature}) vanishes, which implies that the vector potential~(\ref{u3_field}) can be gauged away, $\vec{A} \to \vec{A}' = 0$. This result can be understood as follows. When $\kappa \to 1$, it is straightforward to show that the basis vectors $\ket{\varphi{'}_m} = \hat{S}_1^\dagger (\Omega) \ket{\varphi_m (\Omega)} $ form a representation of the rotation group, $\bra{\varphi{'}_n} \vec{\Lambda} \ket{\varphi{'}_m} = \vec{\sigma}^1_{n m} $. As rotational invariance represents a global symmetry, there cannot exist a gauge field. As a consequence, the Hamiltonian~(\ref{gauge_cov_qp}) decouples into the impurity term, $-B \vec{\nabla}^2$, and the scalar potential, $A^0$. From the physical point of view, the limit  of $\kappa \to 1$ corresponds to the regime where the impurity and the bath  interact only through  an electric potential. 

\begin{figure}
  \centering
  \includegraphics[width=\linewidth]{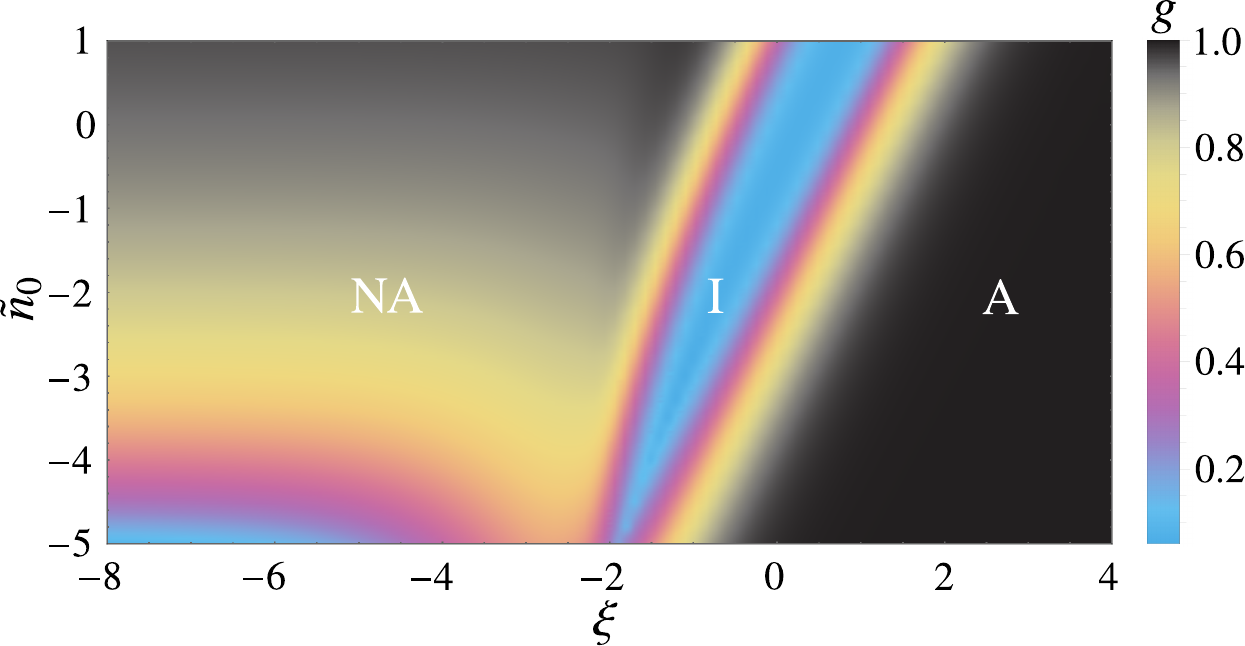}
 \caption{ The monopole charge of the angulon $L=1$ state, $g$, as a function of the dimensionless rotational constant, $\xi = \ln [B/u_0]$, and the dimensionless density, $\tilde{n}_0 =\ln [n_0 (m u_0)^{-3/2}] $. Topological transition from the non-abelian (NA) to the abelian (A) monopole takes place in the angulon instability region (I). See text.}
 \label{tp}
\end{figure}

Away from the instability point $\xi_c$, the monopole charge assumes a finite value. For $\xi < \xi_c$ and for $\xi > \xi_c$, however, the behavior of the gauge field is quite different. In the former regime, the impurity interacts with an effective gauge field which, as we show below, is truly non-abelian. In this regime the interaction couples different internal degrees of freedom of the impurity and the monopole charge takes values in the range of $\sim 0.4 - 0.6$. For $\xi > \xi_c$, on the other hand, the monopole charge is identically one, which corresponds to $\kappa = 0$. In this situation, the monopole gauge field becomes `abelianized,' i.e.\
\be
A_\phi = \cos \theta \, \sigma^1_z\, , \quad A_\theta = 0 \, .
\ee
In other words, for $\xi>\xi_c$ the gauge field can be decomposed into three $U(1)$ gauges fields, $\vec{A} = \vec{A}_- \oplus \vec{A}_0 \oplus \vec{A}_+$. While $\vec{A}_0 = 0$, the $U(1)$  gauge field $\vec{A}_\pm$ is the so-called Dirac monopole field with charge $g_{\pm}=\pm 1$~\cite{Dirac_1931,Wu_Yang_1975}. Thus, in the regime of $\xi > \xi_c$ the components of the state describing the impurity, $\Phi_{\pm}^{\alpha}  (\vec{r})$, interact with the corresponding Dirac monopole without being coupled to each other. The same picture is still valid if we vary the density, cf.\ Fig.~\ref{tp} and the corresponding discussion below.

In order to see that the magnetic vector potential is truly non-abelian for $\xi < \xi_c$, one can argue as follows. Let us assume there exists a gauge transformation that brings the vector potential to the form $\vec{A} = \vec{A}_- \oplus \vec{A}_0 \oplus \vec{A}_+$ which means that $\vec{A} (\Omega)$ is a diagonal matrix for all $\Omega$. Using Eq.~\eqref{curvature}, it is not difficult to see that the integral over the field strength tensor is given by $1/(2 \pi) \int_{S^2} F_{\phi \theta} \, \text{d}\Omega = \text{diag}(-g,0,g)$. That is, the integrated field strengths of $\vec{A}_{-}$, $\vec{A}_0$ and $\vec{A}_{+}$ are given by $-g$, $0$ and $g$, respectively. It is well known, see Ref~\cite{nakahara2003geometry}, that those numbers are the first Chern numbers (topological invariants) of the line bundles associated with $\vec{A}_{-}$, $\vec{A}_0$ and $\vec{A}_{+}$. By definition, they can take only integer values. From the physics perspective, this reflects the well-known fact that $U(1)$ magnetic monopoles are necessarily quantized. Since $g$ lies in the range of $\sim 0.4 - 0.6$, this contradicts the assumption that $\vec{A}$ is abelian and we conclude that our vector potential is truly non-abelian. For non-abelian monopoles the above reasoning fails since in this case the eigenvalues of $1/(2 \pi) \int_{S^2} F_{\phi \theta} \, \text{d}\Omega $ are not topological invariants. The only topological invariant in this situation is the Chern number of the whole bundle, see Ref.~\cite{jiangmini}, which is given by $1/(2 \pi) \int_{S^2} \text{tr}\left( F_{\phi \theta} \right) \, \text{d}\Omega$ and equals zero in our case. The fact that $g$ equals a Chern number in the abelian case also explains why it is constant for all values of the external parameters. On the one hand, it depends continuously on the rotational constant and on the density of the bath and on the other hand, as a Chern number, it can assume only integer values. To change its value, it would have to jump from zero to some other integer value when we vary the system's parameters only a little. Since this is forbidden by continuity, $g$ has to remain constant.

The above discussion reveals that the transition from a non-abelian vector potential with no topological restriction on $g$ for $\xi < \xi_c$ to an abelian vector potential with the topological restriction that $g$ has to be an integer for $\xi > \xi_c$, is also a topological transition of the underlying vector bundle. This topological transition is clearly visible in Fig.~\ref{tp}, where we plot the monopole charge $g$ as a function of both the dimensionless rotational constant, $\xi$, and the dimensionless bath density, $\tilde n_0$. In the non-abelian domain~(NA), the monopole charge takes a range of values between zero and one. On the other hand, for large positive $\xi$, the monopole charge is equal to one with high precision, which corresponds to the abelian region~(A). The topological transition from the non-abelian to abelian monopole takes place across the angulon instability~(I).

In conclusion, we have demonstrated that  a rotating impurity coupled to a many-particle bath (the `angulon quasiparticle')  can be interpreted as a quantum particle on the 2-sphere interacting with a gauge field of a non-abelian magnetic monopole. Recently it has been shown experimentally that molecules immersed in superfluid helium nanodroplets in fact form angulons~\cite{lemeshko2016quasiparticle, YuliaPhysics17, Shepperson16}.  In the particular setting considered here, a superfluid helium droplet manifests  itself as a $U(3)$ non-abelian magnetic monopole in the real space of the molecular impurity.  We demonstrate that the $U(3)$ gauge field vanishes exactly at the angulon instabilities. Furthermore, on one side of the instability the gauge field is truly non-abelian, whereas on the other side the gauge field abelianizes and the internal degrees of freedom of the impurity effectively interact with separate $U(1)$ Dirac monopoles. The abelianization of the gauge field around the instability corresponds to a topological transition of the underlying vector bundle. Since the angulon instabilities have been recently identified in experiment~\cite{Cherepanov}, our results pave the way for the study of topological transitions and related physics using molecules in helium nanodroplets.

\begin{acknowledgments}

We are grateful to G. Bighin, S. Castrignano, S. Meuren, B. Midya, D. Monaco, R. Schmidt, M. Serbyn, and H. Weimer for valuable discussions. E. Y. acknowledges financial support received from the People Programme (Marie Curie Actions) of the European Union's Seventh Framework Programme (FP7/2007-2013) under REA grant agreement No. [291734]. A.~D. acknowledges support from the  European Research Council (ERC) under the European Union's Horizon 2020 research and innovation programme (grant agreement No. 694227). M. L. acknowledges support from the Austrian Science Fund (FWF), under project No. P29902-N27.
\end{acknowledgments}

\bibliography{angulon.bib}

\end{document}